# Linear Rotary Optical Delay Lines

**Maksim Skorobogatiy**

*Department of Engineering Physics, École Polytechnique de Montréal, Québec, Canada*
[*maksim.skorobogatiy@polymtl.ca*](maksim.skorobogatiy@polymtl.ca)

**Abstract:** I present several classes of analytical and semi-analytical solutions for the design of high-speed rotary optical delay lines that use a combination of stationary and rotating curvilinear reflectors. Detailed analysis of four distinct classes of optical delay lines is presented. Particularly, I consider delay lines based on a single rotating reflector, a single rotating reflector and a single stationary reflector, two rotating reflectors, and two rotating reflectors and a single stationary reflector. I demonstrate that in each of these cases it is possible to design an infinite variety of the optical delay lines featuring linear dependence of the optical delay on the rotation angle. This is achieved via optimization of the shapes of rotating and stationary reflector surfaces. Moreover, in the case of two rotating reflectors a convenient spatial separation of the incoming and outgoing beams is possible. For the sake of example, all the blades presented in this paper are chosen to fit into a circle of 10cm diameter and these delay lines feature in excess of 600ps of optical delay.

**References and links**

1. R.E. Beselt, "Large Amplitude High Frequency Optical Delay," Honeywell ASCa, US 7,899,281 (2011)
2. D.C. Edeistein, R. B. Romney and M. Scheuerman, "Rapid programmable 300 ps optical delay scanner and signal averaging system for ultrafast measurements," Rev. Sci. Instrum. 62, 579 (1991)
3. J. Ballif, R. Gianotti, Ph. Chavanne, R.Wälti, and R. P. Salathé, "Rapid and scalable scans at 21 m/s in optical low-coherence reflectometry," Opt. Lett. 22, 757 (1997).
4. J. Szydlo, N. Delachenal, R. Gianotti, R. Walti, H. Bleuler and P.R. Salathe, "Air-turbine driven optical low coherence reflectometry at 28.6-kHz scan repetition rate," Opt. Commun. 154, 1 (1998)
5. G. Lamouchea, M.Dufoura, B. Gauthier a, V.Bartulovicb, M. Hewkoc, and J.-P. Monchalina, "Optical delay line using rotating rhombic prisms," Proc. of SPIE Vol. 6429 64292G-1 (2007)
6. Amy L. Oldenburg, J. Joshua Reynolds, Daniel L. Marks, and Stephen A. Boppart, "Fast-Fourier-domain delay line for *in vivo* optical coherence tomography with a polygonal scanner," Appl. Opt. 42, 4606 (2003).
7. Pei-Lin Hsiung, Xingde Li, Christian Chudoba, Ingmar Hartl, Tony H. Ko, and James G. Fujimoto, "High-speed path-length scanning with a multiple-pass cavity delay line," Appl. Opt. 42, 640 (2003)
8. X. Liu, M. J. Cobb, and X. Li, "Rapid scanning all-reflective optical delay line for real-time optical coherence tomography," Opt. Lett. 29, 80-82 (2004)
9. Kitsakorn Locharoenrat and I-Jen Hsu, "Optical Delay Line for Rapid Scanning Low-Coherence Reflectometer," Intern. J. Inform. Electron. Eng. 2, 904 (2012)
10. C.-L. Wnag, C.-L.Pan, "Scanning Optical Delay Device Having a Helicoid Reflecting Mirror," National Science Council, Taiwan, US 5,907,423 (1999)
11. T.D. Dorney, "Scanning Optical Delay Line Using a Reflective Element Arranged to Rotate," US 7,046,412 (2006)
12. G.J. Kim, Y.S. Jin, S.G. Jeon, J.I. Kim, "Rotary Optical Delay Line," Korea Electrotechnology Research Institute, US 7,453,619 (2008)
13. J. Xu and X.-C. Zhang, "Circular involute stage," Opt. Lett. 29, 2082 (2004)
14. Geun-Ju Kim, Seok-Gy Jeon, Jung-Il Kim, and Yun-Sik Jin, "High speed scanning of terahertz pulse by a rotary optical delay line," Rev. Scien. Instr. 79, 106102 (2008)

## 1. Introduction

Due to relatively low loss of most dry materials in the THz spectral range, and due to strong sensitivity of the material losses to the humidity content, there has been a large interest in using THz waves in non-destructive imaging and process control. In particular, various industrial applications of THz spectroscopy are currently under investigation with potential applications in detection of inorganic inclusions in food products, detection of dangerous chemicals disguised by packaging, detection of defects in packaged electronic chips, monitoring of water content in paper and textile fabrication lines, to name just a few.

In our research group we pursue design of THz spectroscopy and imaging systems that are based on photoconductive antenna sources and detectors. Ideally, such systems should feature low noise and high acquisition rates. Photoconductive antenna-based systems have gained considerable attention due to wide commercial availability of the broadband photoconductive antennas, as well as due to availability of relatively cheap fiber-based femtosecond lasers that are used to pump photoconductive antennas. Very high signal-to-noise ratios of up to 50-60dB can be achieved in such systems when using lock-in amplifiers together with high chopping rates. Moreover, low efficiencies of a single dipole-type THz antenna can be compensated by using high numerical aperture antennas or the interdigitated antenna arrays that can produce as much as ~100μW of THz power with ~1W femtosecond pump laser.

For rapid terahertz imaging and real-time process monitoring, a high scanning rate optical delay line is required for efficient sampling of the terahertz pulses. In addition to high scanning rate, one also desires high resolution, large amplitude of scanning, high repeatability, and easy encoding/decoding of the optical delay value. Currently, majority of the THz systems use delay lines based on linear micropositioning stages with mounted retroreflectors. The popularity of this solution is related to the maturity of the micropositioning technology and its relatively low price. Moreover, optical delay produced by such delay lines depends linearly on the stage displacement, which greatly simplifies encoding and readout of the optical delay value. Spectral resolution $\Delta \nu$ of a THz spectroscopy setup is related to the maximal usable optical delay $t_{max}$ and is given by $\Delta \nu \approx 1/t_{max}$. While even a small $L = 15\,cm$-long linear micropositioning stage can provide as much as $t_{max} = 2 \cdot L \cdot c = 1000\,ps$ of optical delay, in practice, the maximal usable delay is typically limited to $t_{max} \sim 200\,ps - 500\,ps$. This is due to appearance of various parasitic echoes and perturbations in the pulse shape incurred because of multiple reflections of the THz beam on the structural elements of a THz setup. This limits resolution of a standard THz setup to about 2GHz. Moreover, the bandwidth of a THz setup is limited by the Nyquist frequency $\nu_{max} \approx 1/(2 \cdot \Delta t)$, where $\Delta t$ is a resolution of the optical delay line. Typically, one requires that Nyquist frequency is several times larger (say 2 times) than the THz bandwidth of interest (~2.5THz), from which it follows the estimate for the delay stage resolution of $\Delta t < 100\,fs$. This, in turn, sets a limit on the accuracy of the linear micropositioning stage that should be superior to the $\Delta L < \Delta t \cdot c/2 \approx 15\,\mu m$.

One of the challenges associated with the use of linear micropositioning stages as optical delay lines is their limited scanning speed. The maximal displacement speed of a linear stage with displacement accuracy superior to $15\,\mu m$ and displacement range of $15\,cm$ typically does not exceed $1\,m/s$. That means that even in the most optimal case the repetition rate of such a delay line (and, hence, acquisition rate of a THz system) will not exceed $\sim 5\,Hz$.

Low acquisition rates set by the use of linear micropositioning stages lead to investigation of various other delay line architectures that can provide higher scanning rates. One of the enhancements on the linear positioning stage was described in [1], where the authors used a rotary to linear mechanical coupling device that allowed them to displace the retroreflector along a linear path, while using fast rotary actuation. A system capable of 100s of ps of optical delay and ~50Hz scanning rate was suggested. Unfortunately, resultant optical delay depends in a non-linear manner on the motor rotation angle, thus, necessitating the use of an optical encoder for precise readout of the instantaneous optical delay. A galvanometer-based oscillating optical delay line [2] was proposed a decade ago for high speed scanning of THz pulses. In this delay line, one of the two mirrors in the optical cavity oscillates with a fixed amplitude and frequency. This system was capable of achieving 300ps optical delay with 30Hz scanning rate. However, the position and velocity of the mirror in this setup changes sinusoidally, so the optical delay did not change linearly in time. Then, a rotating cube [3,4] (or a prism [5]) was suggested. In such systems the light is first passed through the rotating cube, then reflected from the stationary mirror and, finally, passed back through the same cube. In this arrangement, the reflected beam follows the same path as the incident beam, therefore requiring separation of the two at the output of the delay line. Scanning rates as high as 380Hz were demonstrated with optical delay in 100s of ps. Placing the cube on the air turbine can further improve the scanning speed by almost an order of magnitude. Unfortunately, the resultant optical delay is highly nonlinear with respect to the rotation angle of a cube, therefore, requiring an optical encoder for precise readout of the optical delay. Moreover, when using larger cubes, group velocity dispersion of the cube material can become important due to long propagation length inside of the cube. This may lead to a femtosecond pulse broadening and, hence, lower efficiencies of THz generation and detection. In [6] a Fourier-domain delay line with a polygonal mirror array was presented and high scanning rates of 4kHz were demonstrated. The design profited from commercial polygonal mirror arrays capable of 15kHz scanning rates, the main disadvantage of this setup was a relatively low optical delay of 10ps, which is adequate for optical coherence tomography applications, but not for THz spectroscopy. Additionally, optical delay depends in a non-linear manner on the rotation angle of the mirror array. Other notable optical delay lines mostly constructed for OCT applications, and, hence characterized by relatively small optical delays (<20ps), and high scanning rates (several kHz) are multiple-pass cavity delay line [7] as well as various combinations of curved or straight mirrors and a scanning mirror [8,9].

Recently, several linear rotary optical delay lines based on the combination of rotating and stationary curvilinear reflectors have been demonstrated [10-14]. The main advantages of these delay lines include : linear dependence of the optical delay on the reflector rotation angle, large values of the optical delay (100s-1000s of ps), high scanning rates (100s of Hz), and relative ease of their fabrication using CNC machines. When using rotary delay lines that feature linear dependence of the optical delay on the reflector rotation angle, one can, in principle, forgo complex optical encoders if the rotation rate is precisely controlled. Moreover, optical delay provided by such systems is proportional to the reflector size. As we will see later in the paper, when using relatively small systems that fit into a 10cm-diameter circle, optical delay in excess of 600ps can be achieved, which is comparable to the optical delay generated by a standard 10cm-long linear micropositioning stage with a retroreflector. Finally, the basic scanning rate of a rotary delay line is typically in the 50Hz range when standard motors are used at 3000RPM. By placing several reflectors on the same rotating base, one can increase the acquisition speed by an order of magnitude. Thus, in [14] the scanning rates of 400Hz and the total delay of 140ps have been demonstrated when using six reflectors, all fitting into an 8cm diameter circle. At this point, it is important to note that although the high scan rates are typically desired, an increase in speed also requires a commensurate increase in optical power to maintain the same signal to noise ratio.

Rotary delay lines require careful design of the shapes of their reflector surfaces. The main constraint in the design of an optical delay line is the demand that incoming and outgoing light paths stay invariant during reflector rotation. Moreover, it is also desirable that the resultant optical delay is linear with respect to the reflector rotation angle. So far, designs available in the literature only present several particular examples of the reflector shapes, while general theory behind rotary delay line design is still missing.

The main goal of this work is to develop a comprehensive theory for the design of rotary delay lines. Particularly, I consider theoretical formulations as well as analytical and semi-analytical solutions for four general classes of the rotary optical delay lines. Delay lines under consideration are comprised of the following components: a single rotating reflector, a single rotating reflector and a single stationary reflector, two rotating reflectors, and, finally, two rotating reflectors and a single stationary reflector. Remarkably, in most cases it is possible to design an infinite variety of the linear optical delay lines via optimization of shapes of the rotating and stationary reflector surfaces. Moreover, in the case of two rotating reflectors a convenient spatial separation of the incoming and outgoing beams is possible. In the case of a single rotating reflector spatial separation of the incoming and outgoing beams is also possible via introduction of the rectangular V groove into the reflector edge. For the sake of example, all the blades presented in this paper are chosen to fit into a circle of 10cm diameter and these delay lines feature in excess of 600ps of optical delay. This is comparable to the optical delay generated by a standard 10cm-long linear micropositioning stage with a retroreflector.

## 2. Optical delay line using a single rotating blade

Example of a rotary optical delay line with one rotating reflector is presented in Fig. 1. There, a curvilinear reflector (which I call a blade in the rest of the paper) is rotating around a fixed axis positioned at the origin. The axis of rotation is perpendicular to the blade surface. Light beam is arriving parallel to the OX axis and is reflected back by the blade edge along the same path. The light beam is displaced along the OY axis by $R_i$. In this simplest implementation that was reported in [1], the rotating blade stays perpendicular to the incoming beam regardless of the blade rotation angle.

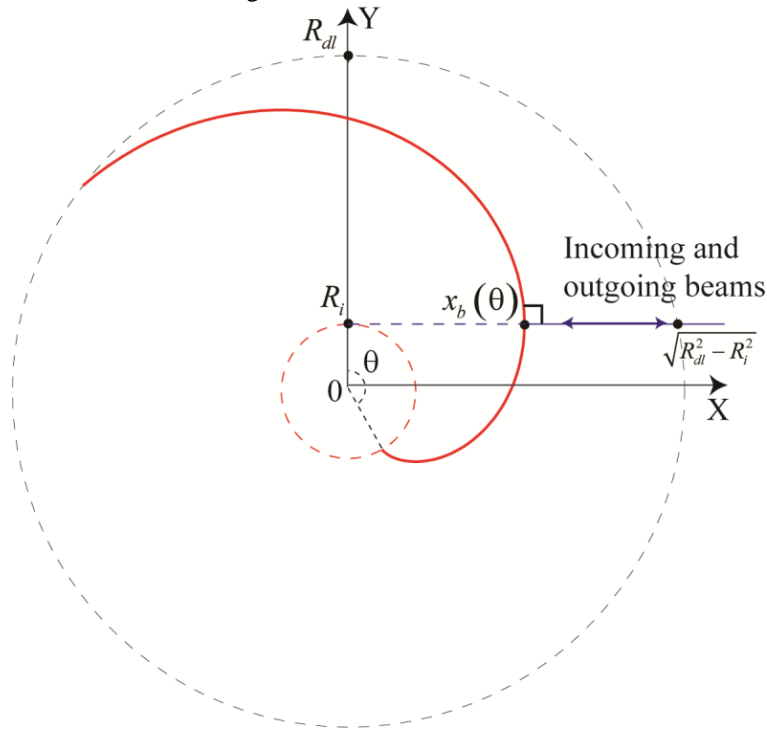

Fig. 1. Schematic of a rotary optical delay line featuring a single rotating blade. By design, the blade edge is always perpendicular to the incoming light beam at all rotation angles. Light beam is arriving parallel to the OX axis and is reflected back by the blade along the same path.

I will now demonstrate that an analytical solution exists for the shape of a rotating blade, which results in the blade being always perpendicular to the incoming light beam at all rotation angles. Moreover, a corresponding delay line will have a linear dependence of the optical delay on the blade rotation angle. In what follows, I define $\theta$ to be the blade rotation angle assuming a clockwise rotation of the blade from the OY axis. When the blade rotates, the point of its contact with the light beam will only change its position along the OX axis according to $x_b(\theta)$, while staying constant at $R_i$ along the OY axis.

First I show that if the functional form of the position of the light reflection point $x_b(\theta)$ on the blade rotation angle $\theta$ is specified, then, the shape of the blade is mostly specified. In what follow I suppose that the blade rotation angle is defined in the interval $\theta \in [0, \theta_{max}]$. I assume that at $\theta = 0$ the shape of the blade in cylindrical coordinate system is given by:

$$x(\varphi) = r(\varphi)\cos(\varphi)$$
$$y(\varphi) = r(\varphi)\sin(\varphi)$$
. (2)

As the blade rotates clockwise by angle $\theta$, its shape can be expressed as:

$$x(\varphi) = r(\varphi)\cos(\varphi - \theta)$$
$$y(\varphi) = r(\varphi)\sin(\varphi - \theta)$$
. (3)

For a given angle of rotation $\theta$, the shape of the blade described by (3) should contain a light reflection point having Cartesian coordinates $(x_b(\theta), R_i)$. From (3) I can then find the parametrical representation of the blade curve in terms of the angle of blade rotation as:

$$\begin{cases} x(\varphi(\theta)) = x_b(\theta) \\ y(\varphi(\theta)) = R_i \end{cases} \Rightarrow \begin{cases} \tan(\varphi(\theta) - \theta) = \dfrac{y}{x} = \dfrac{R_i}{x_b(\theta)} \\ r(\varphi(\theta)) = \sqrt{x^2 + y^2} = \sqrt{x_b(\theta)^2 + R_i^2} \end{cases}, (4)$$

from which it follows that the blade shape at $\theta = 0$ is given by the following parametric expression:

$$\varphi = t + \arctan\left(\dfrac{R_i}{x_b(t)}\right) + \pi n$$

$$r(\varphi) = \sqrt{x_b(t)^2 + R_i^2} \quad , (5)$$

$$t \in [0, \theta_{max}], \, n \in \mathbb{Z}$$

where I have deliberately introduced a new parameter $t$ that should not be confused with a blade rotation angle $\theta$. Note that parameter $t$ is defined on the same interval $[0, \theta_{max}]$ as the blade rotation angle. Finally, in Cartesian coordinates the blade shape is defined by substitution of (5) into (2).

## 2.1 A blade perpendicular to the incoming light

As derived in the appendix, the inclination angle of the blade with respect to the direction of an incoming light (OX axis) is:

$$\tan(\alpha_b) = \dfrac{x_b(\theta)}{\partial x_b(\theta)/\partial\theta - R_i}. (6)$$

From this it follows that the choice of:

$$x_b(\theta) = x_0 + R_i \cdot \theta, (7)$$

results in the blade that is always perpendicular to the incoming light ($\alpha_b = \pi/2$) independently of the rotation angle. Moreover, optical delay provided by such a blade has a linear dependence on the blade rotation angle, with the value of the optical delay given by:

$$\Delta T = \frac{2}{c} \cdot \left( x_b(\theta) - x_b(0) \right) = \frac{2 R_i}{c} \cdot \theta \; ; \; \theta \in [0, \theta_{max}] . \quad (8)$$

Throughout the paper I suppose that the value of optical delay is zero for a zero rotation angle.

**2.2 Maximal optical delay and multiple blade designs**

Now, I derive the maximal optical delay achievable by the delay line featuring the blade perpendicular to the incident light. As a constraint I demand that such a blade should be completely inscribed into a circle of radius $R_{dl}$ (see Fig. 2(a)). The maximal optical delay is achieved when the light reflection point spans all the possible values $x_b(\theta) \in \left[ 0, \sqrt{R_{dl}^2 - R_i^2} \right]$ for the corresponding rotation angles $\theta \in [0, \theta_{max}]$. Additionally I require that $0 < \theta_{max} < 2\pi$ in order to avoid blocking the incoming light with a part of the same blade. From (7) I find:

$$x_b(0) = 0 \Rightarrow x_0 = 0$$

$$x_b(\theta_{max}) = \sqrt{R_{dl}^2 - R_i^2} \Rightarrow R_i = \frac{R_{dl}}{\sqrt{\theta_{max}^2 + 1}} . \quad (8.1)$$

Therefore, from (8) it follows that the maximal total delay is given by:

$$\Delta T_{max} = \frac{2 R_{dl}}{c} \cdot \frac{\theta_{max}}{\sqrt{\theta_{max}^2 + 1}} . \quad (8.2)$$

From (8.2) it follows that there is a fundamental limit to the maximal achievable optical delay produced by a rotary delay line of radius $R_{dl}$ which is $\Delta T_{max} = 1.975 \cdot R_{dl}/c$, when $\theta_{max} = 2\pi$. For example, in Fig. 2(a),(b) we present the case of a blade inscribed into a circle of radius $R_{dl} = 5 cm$ and having $\theta_{max} \approx 1.56 \cdot \pi$, which results in a somewhat suboptimal delay $\Delta T \approx 326 ps < \Delta T_{max} \approx 329 ps$.

It is also interesting to note from (8.1) that multiple blades ($p$ blades) can be accommodated in the same rotary delay line if $\theta_{max}$ is a simple fraction of $2\pi$, that is, $\theta_{max} = 2\pi/p$ (see Figs. 2©,(d)). This sets the relation between the two radii as $R_i = R_{dl}/\sqrt{1 + (2\pi/p)^2}$ and the maximal achievable optical delay becomes:

$$\Delta T_{max} = \frac{2 R_{dl}}{c} \cdot \frac{2\pi}{\sqrt{p^2 + (2\pi)^2}} . \quad (8.3)$$

Using multiple blades ($p$ blades) allows to increase the scanning rate by a factor of $p$. Moreover, as follows from (8.3), if the number of blades is less than 6, this increase in the scanning rate comes only with a slight reduction in the maximal optical delay.

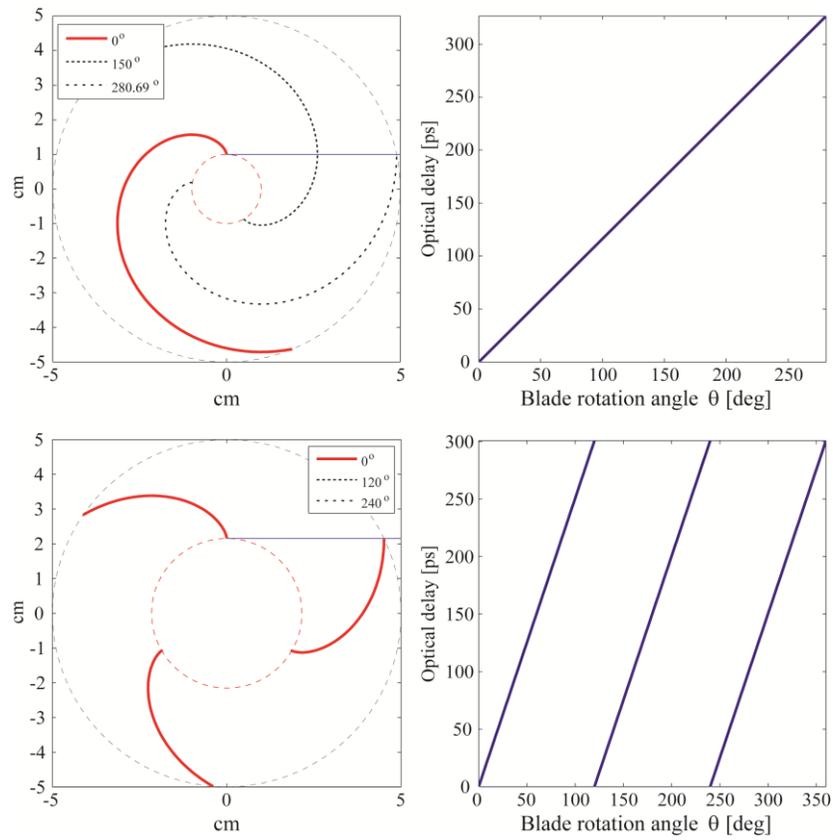

Fig. 2. (a) Schematic of a rotary optical delay line featuring a single rotating blade. Blade positions at various rotation angles are also presented as dotted lines. (b) Corresponding optical delay. (c) Schematic of a rotary delay line with 3 blades and (b) its corresponding optical delay.

## 2.3 Spatial separation of the incoming and outgoing beams

Introduction of the rectangular V groove structure into the reflector edge (see Fig 2.1) allows separation of the incoming and outgoing beams. Particularly, the V groove has a $90^o$ angle and it is inscribed into the blade perpendicular to the reflector curve shown in Fig. 2. In fact, the V groove serves as a retroreflector for the incoming light beam, thus separating the incoming and the outgoing beams in space. Finally, the V groove is positioned in such a way that the incoming and outgoing beams see the blade shape described by (5).

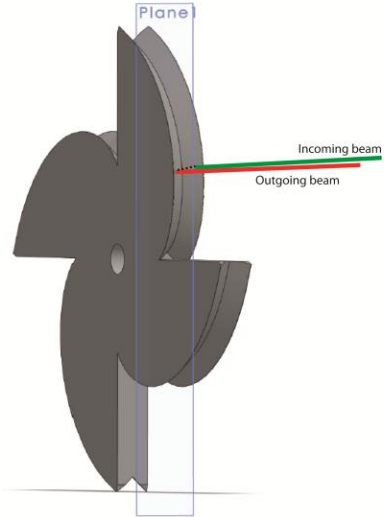

Fig. 2.1. Schematic of a rotary optical delay line that is capable of separating incoming and outgoing beams. This shape of a linear delay line is similar to the one presented in Figs. 1,2, except it features an additional rectangular V groove inscribed into the blade edge. The V groove serves as a retroreflector for the incoming beam.

## 3. Optical delay line using a single rotating blade and a single stationary reflector

In this section I consider several example of rotary optical delay lines that comprise one inclined blade and one stationary reflector (see Figs. 3(a), 4(a)). As before, I consider a curvilinear reflector (a blade) rotating around a fixed axis. The light beam contacts the blade at $x_b(\theta)$ along the OX axis. The light beam is arriving parallel to the OX axis and it is displaced by $R_i$ along the OY direction. This time, however, the blade surface is not perpendicular to the incoming beam. In fact, I will distinguish two cases.

In the first case (see Fig. 3(a)), the blade edge makes a constant angle $\alpha_b$ with respect to the direction of the incoming beam (OX axis). Therefore, I only need to use a simple planar stationary reflector to send the beam back along its arrival path. I will demonstrate that in this simple case I can only realise a nonlinear optical delay that changes exponentially with the rotation angle.

In the second case, I allow the angle of the blade edge with respect to the beam direction to vary with the rotation angle (see Fig. 4(a)). In this more general case I need to use a curvilinear stationary reflector to send the light back along its arrival path. At the same time, I show that in this more complicated case there exist an infinite number of analytical solutions for the blade shape that result in the linear dependence of the optical delay with respect to the rotation angle. In one notable case, the blade shape is a simple straight line, while the shape of a stationary reflector is curvilinear (see Fig. 4(a)); this case is of particular interest for practical realisation of the linear rotary optical delay line.

### 3.1 A blade having constant inclination angle with respect to the incoming light

I now consider the case when the blade forms a fixed angle $\alpha_b \neq \pi/2$ with the direction of an incoming light. In this case one can realise the optical delay line by using an additional stationary planar reflector to direct the light back along its original path of arrival (see Fig. 3(a)). For that, the stationary reflector should be perpendicular to the light reflected by the blade. From (6) I can now find the displacement of the light reflection point as a function of the rotation angle:

$$\tan(\alpha_b) = \frac{x_b(\theta)}{\partial x_b(\theta)/\partial \theta - R_i}. \quad (9)$$

Solution of equation (6) satisfies the following differential equation:

$$\frac{\partial x_b(\theta)}{\partial \theta} - \cot(\alpha_b) \cdot x_b(\theta) - R_i = 0, \quad (10)$$

which allows analytical solution of the following form:

$$x_b(\theta) = x_0 \cdot \exp(\cot(\alpha_b) \cdot \theta) - \tan(\alpha_b) \cdot R_i. \quad (11)$$

In order to calculate optical delay provided by the rotary delay line I have to first find the coordinates of a light reflection point at the surface of a stationary reflector. If the blade forms a fixed angle $\alpha_b$ with the OX axis, then, the reflected light forms an angle $\alpha_r = 2\alpha_b - \pi$ with the OX axis. Therefore, light reflected by the blade travels along the line:

$$y = R_i + (x - x_b(\theta)) \cdot \tan(2\alpha_b - \pi). \quad (12)$$

In the following, I denote $L(\theta)$ to be the distance travelled by light between the blade and the stationary reflector. From (12) it then follows that coordinates of the light reflection point located on the surface of a stationary reflector are given by:

$$\begin{aligned} x_{sr}(\theta) &= x_b(\theta) - L(\theta)\cos(2\alpha_b) \\ y_{sr}(\theta) &= R_i - L(\theta)\sin(2\alpha_b) \end{aligned}. \quad (13)$$

At the same time, the stationary reflector should be perpendicular to the light reflected by the blade by making the $2\alpha_b - \pi/2$ angle with the OX axis. In other words, the tangent to the stationary reflector surface should take a constant value as follows:

$$\frac{\partial y_{sr}(\theta)/\partial \theta}{\partial x_{sr}(\theta)/\partial \theta} = \tan(2\alpha_b - \pi/2) \Rightarrow \frac{\partial L(\theta)}{\partial \theta} = \frac{\partial x_b(\theta)}{\partial \theta} \cdot \cos(2\alpha_b). \quad (14)$$

The differential equation with respect to $L(\theta)$ in (14) was derived using definitions (13), which has a particularly simple solution:

$$L(\theta) = L_0 + x_b(\theta) \cdot \cos(2\alpha_b), \quad (15)$$

where $L_0$ is some constant. Finally, I conclude that the optical delay provided by such a delay line has a non-linear dependence as a function of the rotation angle, with the absolute value of the optical delay given by:

$$\begin{aligned} \Delta T &= \frac{2}{c} \cdot (x_b(\theta) - L(\theta) - x_b(0) + L(0)) \\ &= \frac{4}{c} \cdot \sin^2(\alpha_b) \cdot (x_b(\theta) - x_b(0)) \end{aligned}. \quad (16)$$

By comparing the optical delays produced by the rotary delay lines that use perpendicular blade (8) and inclined blade (16), I notice that the letter can give twice as large optical delay than the former, which is due to additional path travelled by the light between the rotating blade and the stationary reflector.

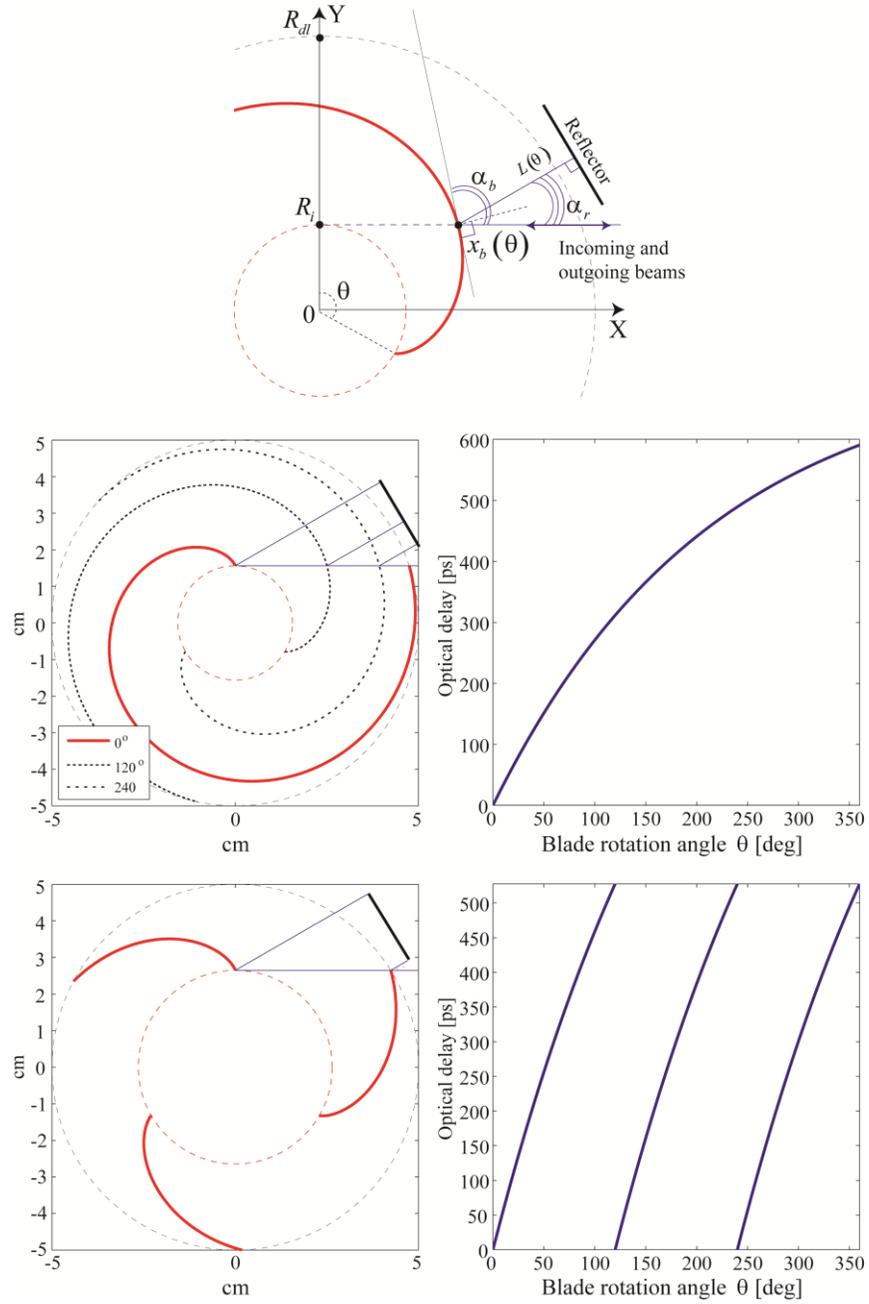

Fig. 3. (a) Schematic of a rotary optical delay line featuring a rotating blade that forms a fixed inclination angle with the direction of incoming light and a linear stationary reflector. Blade positions at various rotation angles are also presented as dotted lines. (c) Schematic of a rotary optical delay line featuring 3 rotating blades. (d) Optical delay for a three-blade system.

### 3.1.1 Maximal optical delay and multiple blade designs

Now, I derive the maximal optical delay achievable by a delay line featuring multiple blades. Each of the blades is inclined at a constant angle with respect to the direction of the incident light. As before, I demand that such a blade is completely inscribed into the circle of radius $R_{dl}$. The maximal total optical delay is achieved when the light reflection point at the blade surface spans all the possible values $x_b(\theta) \in \left[0, \sqrt{R_{dl}^2 - R_i^2}\right]$, while $\theta \in [0, 2\pi/p]$, where $p$ is the number of blades in the design. As $x_b(\theta)$ is a monotonous function of the rotation angle, then using (11) I then find:

$$x_b(0) = 0 \Rightarrow x_0 = \tan(\alpha_b) \cdot R_i$$

$$x_b(2\pi/p) = R_{dl} \Rightarrow R_i = \frac{R_{dl}}{\sqrt{1 + \tan^2(\alpha_b) \cdot \left(\exp\left(\cot(\alpha_b) \cdot \frac{2\pi}{p}\right) - 1\right)^2}}. \quad (17)$$

From this I find the maximal optical delay of a rotary delay line with $p$ blades:

$$\Delta T_{max} = \frac{4 R_{dl}}{c} \cdot \sin^2(\alpha_b) \cdot \sqrt{1 - \frac{R_i^2}{R_{dl}^2}}. \quad (18)$$

As an example, consider $R_{dl} = 5\,cm$, $\alpha_b = \pi/2 + \pi/12$, (see Fig. 3(b),(c)) which results in $\Delta T_{max} \approx 591\,ps$. In the case $\alpha_b > \pi/2$, and in the limit $|\tan(\alpha_b)| \ll 2\pi/p$, the maximal optical delay can be simply expressed as:

$$\Delta T_{max} = \frac{4 R_{dl}}{c} \cdot \sin^3(\alpha_b). \quad (19)$$

From (18), (19) I note that optical delay of a rotary delay line with inclined blade can be designed to be up to twice as large as that of a delay line with a perpendicular blade. Also, for the same overall blade size, depending on the blade inclination angle the maximal optical delay can be varied from zero to its maximal value given by (19). This allows designing both the high resolution delay lines with small optical delays when $0.71 \cdot \pi \ll \alpha_b < \pi$, as well as lower resolution delay lines with high optical delay when $\pi/2 < \alpha_b \ll 0.71 \cdot \pi$.

### 3.2 A blade having changing inclination angle with respect to the incoming light

As I have demonstrated in the previous section, using inclined blades allows exploration of a broad design space for the rotary optical delay lines. A limiting feature of the design presented in the previous section is a highly nonlinear dependence of the optical delay on the blade rotation angle $\theta$. In this section I consider a more general case when the blade inclination angle $\alpha_b(\theta)$ with respect to the direction of incoming light can change with the rotation angle. In this case, one has to use a curvilinear stationary reflector in order to reflect the light back along its original path of arrival. Additionally, I demonstrate that there exist an infinite number of the delay line designs that feature linear dependence of the optical delay on the rotation angle. Notably, there exists a particularly simple design for the linear delay line with the rotating blade in the form of a simple planar surface (see Fig 4(a)).

As before, I consider a rotating blade with the shape that is defined by the $\left(x_b(\theta), R_i\right)$ coordinates of the light reflection point on the blade surface. For the blade inclination angle $\alpha_b(\theta)$ with respect to the direction of incoming light I use (6):

$$\tan\left(\alpha_b(\theta)\right) = \frac{x_b(\theta)}{\partial x_b(\theta)/\partial \theta - R_i}. \quad (20)$$

In order to calculate optical delay provided by the rotary delay line I have to first find the coordinates of the light reflection point on the surface of a stationary reflector. If the blade forms an angle $\alpha_b(\theta)$ with the OX axis, then the reflected light forms an angle $\alpha_r(\theta) = 2\alpha_b(\theta) - \pi$ with the OX axis. Therefore, light reflected from the blade travels along the line:

$$y = R_i + \left(x - x_b(\theta)\right) \cdot \tan\left(2\alpha_b(\theta) - \pi\right). \quad (21)$$

In the following, I denote $L(\theta)$ to be the distance travelled by light between the blade and the stationary reflector. From (21) it also follows that coordinates of the second light reflection point located on the surface of a stationary reflector are given by:

$$\begin{aligned} x_{sr}(\theta) &= x_b(\theta) - L(\theta)\cos\left(2\alpha_b(\theta)\right) \\ y_{sr}(\theta) &= R_i - L(\theta)\sin\left(2\alpha_b(\theta)\right) \end{aligned}. \quad (22)$$

At the same time, the stationary reflector should be perpendicular to the light reflected by the blade by making the $2\alpha_b(\theta) - \pi/2$ angle with the OX axis. In other words, the tangent to the stationary reflector surface should take a specific value as follows:

$$\frac{\partial y_{sr}(\theta)/\partial \theta}{\partial x_{sr}(\theta)/\partial \theta} = \tan\left(2\alpha_b(\theta) - \pi/2\right). \quad (23)$$

By substituting (20) and (22) into (23) I find that the distance between the blade and the stationary reflector satisfies the following differential equation:

$$\frac{\partial L(\theta)}{\partial \theta} = \frac{\partial x_b(\theta)}{\partial \theta}\cos\left(2\alpha_b(\theta)\right). \quad (24)$$

If $x_b(\theta)$ is specified, then, using expression (20) for the blade inclination angle, as well as identity $\cos(2\alpha) = \left(1 - \tan^2(\alpha)\right)/\left(1 + \tan^2(\alpha)\right)$, I can solve (24) by integration:

$$L(\theta) = L_0 + \int_0^{\theta_{\max}} d\theta \frac{\partial x_b(\theta)}{\partial \theta}\left[1 - \left(\frac{x_b(\theta)}{\partial x_b(\theta)/\partial \theta - R_i}\right)^2\right]\left[1 + \left(\frac{x_b(\theta)}{\partial x_b(\theta)/\partial \theta - R_i}\right)^2\right]^{-1}. \quad (25)$$

### 3.2.1 Rotary delay line with linear dependence of the optical delay on rotation angle

I now investigate designing of rotary delay line featuring linear dependence of the optical delay on the blade rotation angle. In this case, the blade shape defined by $x_b(\theta)$ should have a very specific form, which I find by requiring that the total length of the light path depends linearly on the blade rotation angle:

$$x_b(\theta) - L(\theta) = \beta \cdot \theta - L_0, \quad (26)$$

where $L_0$ is some constant. From (26) I can then express $L(\theta)$ in terms of $x_b(\theta)$, and after substitution into (24) I find the following differential equation:

$$\frac{\partial x_b(\theta)}{\partial \theta} = \frac{\beta}{2}\left(1 + \frac{1}{\tan^2(\alpha_b(\theta))}\right). \quad (27)$$

Finally, after substituting expression (20) for the blade inclination angle into (27) I get the following differential equation for the blade shape:

$$\frac{\partial x_b(\theta)}{\partial \theta} = \frac{\beta}{2}\left(1 + \frac{(\partial x_b(\theta)/\partial\theta - R_i)^2}{x_b^2(\theta)}\right). \quad (28)$$

Remarkably, a highly nonlinear differential equation (28) can be solved semi-analytically. In particular, (28) can be first rewritten as a problem of finding the two roots of the second order polynomial with respect to the derivative of $x_b(\theta)$:

$$\left(\frac{\partial x_b(\theta)}{\partial \theta}\right)^2 - 2\left(R_i + \frac{x_b^2(\theta)}{\beta}\right)\frac{\partial x_b(\theta)}{\partial \theta} + \left(x_b^2(\theta) + R_i^2\right) = 0, \quad (29)$$

which has the following solutions:

$$\frac{\partial x_b(\theta)}{\partial \theta} = R_i + \frac{x_b^2(\theta)}{\beta} \pm \sqrt{\left(\frac{x_b^2(\theta)}{\beta}\right)^2 + \left(\frac{2R_i}{\beta} - 1\right)x_b^2(\theta)}. \quad (30)$$

Finally, (30) can be solved by integration with respect to $x_b$. The resulting expression gives dependence of the blade rotation angle on the value of the light reflection point displacement:

$$\theta(x_b) = C_0 + \int_{x_b^{\min}}^{x_b} dx \left[R_i + \frac{x^2}{\beta} \pm \sqrt{\left(\frac{x^2}{\beta}\right)^2 + \left(\frac{2R_i}{\beta} - 1\right)x^2}\right]^{-1}, \quad (31)$$

where $C_0$ is some constant. Note that solution (30) describes a continuum of all the possible designs for the linear rotary delay line parametrized by two parameters $R_i$ and $\beta$.

### 3.2.2 Linear delay line with a straight blade

There exists a particular simple solution of (28) that describes a linear blade and a curvilinear reflector that together comprise a linear rotary delay line (see Fig. 4(a)). Particularly, I consider $\beta = 2R_i$ in the integral solution (31). While, in principle, expression (31) defines two solutions depending on the choice of a sign, I find that one of the solutions is in the form of $x_b = x_0 + R_i\theta$ that defines a blade which is always perpendicular to the incoming light (see Section 2.1). I am, therefore, interested in the second solution of (31) that has the following form:

$$\theta(x_b) = C_0 + \int_{x_b^{\min}}^{x_b} dx \left[R_i + \frac{x^2}{R_i}\right]^{-1} = \theta_0 + \arctan\left(\frac{x_b}{R_i}\right). \quad (32)$$

From (32) I, finally find the blade parameters:

$$x_b(\theta) = R_i \tan(\theta - \theta_0)$$
$$L(\theta) = L_0 + R_i \tan(\theta - \theta_0) - 2R_i \cdot \theta \quad . (33)$$

From (4) I confirm that the blade shape is a straight line that forms an angle $\pi/2 + \theta_0 - \theta$ with the OX axis. The value of the optical delay produced by such a delay line is then:

$$\Delta T = \frac{2}{c} \cdot (x_b(\theta) - L(\theta) - x_b(0) + L(0)) = \frac{4R_i}{c}\theta \quad . (34)$$

Finally, I note that the optical delay produced by the straight blade and a stationary reflector (34) is exactly twice as large as the one produced by the blade that is always perpendicular to the incoming light (8). This is simply due to additional path travelled by the light between the blade and the stationary reflector.

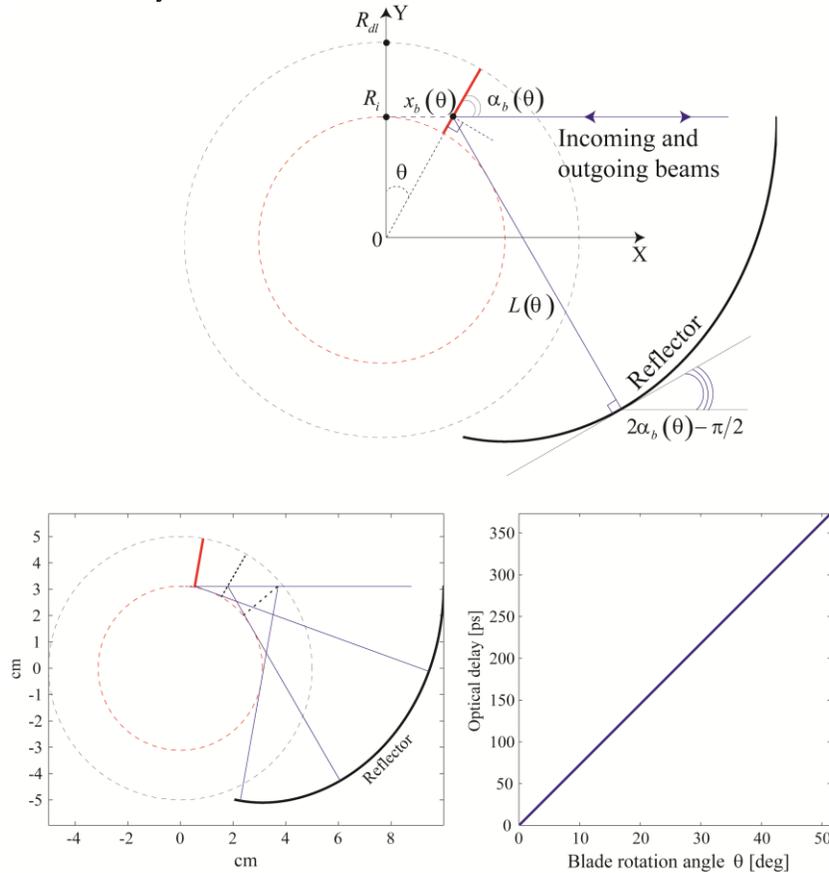

Fig. 4. (a) Schematic of a linear rotary optical delay line featuring a straight rotating blade and a stationary curvilinear reflector. (b) Rotary stage design that can accommodate $p = 7$ straight blades. Blade positions at various rotation angles are presented as dotted lines. (c) Corresponding optical delay.

### 3.2.3 Maximal optical delay and multiple blade designs

Consideration of this section are virtually identical to those of Section 2.2, therefore I only present the final results. First, without the loss of generality I assume that $\theta_0 = 0$ in (33) which corresponds to a straight blade that is perpendicular to the direction of the incoming light for a zero rotation angle. By demanding that the rotating blade is completely inscribed into a circle of radius $R_{dl}$ I then conclude that the maximal optical delay is achieved when the light reflection point spans all the possible values $x_b(\theta) \in \left[0, \sqrt{R_{dl}^2 - R_i^2}\right]$ with the corresponding rotation angles $\theta \in [0, \theta_{max}]$, $\theta_{max} = \arccos(R_i/R_{dl})$. The total delay achieved by this stage is therefore given by:

$$\Delta T = \frac{4 R_i}{c} \cdot \arccos\left(\frac{R_i}{R_{dl}}\right), \quad (35)$$

while the maximal delay is achieved for an optimal choice of $R_i$:

$$\Delta T_{max} = \max_{R_i}(\Delta T) \approx \frac{2.244 \cdot R_{dl}}{c}. \quad (36)$$

$$R_i \approx 0.652 \cdot R_{dl}$$

For example, for a blade radius $R_{dl} = 5 cm$ this results in $\Delta T_{max} = 374 ps$. Note that a rotary delay line with the maximal optical delay given by (36) can accommodate up to 7 non overlapping blades. When distributed uniformly (anglewise) there will be small gaps between the blades. In general, $p > 4$ blades without gaps between them can be accommodated into the rotary delay line if the individual blade maximal rotation angle $\theta_{max}$ is a simple fraction of $2\pi$. This sets the relation between the two radii as $R_i = R_{dl} \cdot \cos(2\pi/p)$ and the maximal achievable optical delay for a delay line with $p$ blades becomes:

$$\Delta T^p = \frac{4 R_{dl}}{c} \cdot \left(\frac{2\pi}{p}\right) \cdot \cos\left(\frac{2\pi}{p}\right). \quad (37)$$

The maximal delay is achieved when $p = 7$ blades are used and it is given by:

$$\Delta T_{max}^p = \max_{p=7}(\Delta T^p) \approx \frac{2.239 \cdot R_{dl}}{c}. \quad (38)$$

Using multiple blades ($p$ blades) allows to increase the scanning rate by a factor of $p$. Moreover, as follows from (37), if the number of blades is less than 15, this increase in the scanning rate comes only with a minimal reduction in the maximal optical delay.

In Fig. 4(b),(c) I present the case of a linear delay line designed to accommodate $p = 7$ straight blades. To simplify the presentation, only one blade is shown in the figure. The parameters of this delay line are chosen as follows $R_{dl} = 5 cm$, $R_i = R_{dl} \cdot \cos(2\pi/7) \approx 3.1\ cm$, thus resulting in a maximal delay of $\Delta T = 374 ps$.

## 4. Linear rotary optical delay line using two rotating blades

Sometimes, it is of interest to separate in space the incoming and outgoing light beams. For optical delay line applications it is, however, important that the line of path of the outgoing light beam is invariable in time. One of such systems is presented in Fig. 5. There, I consider a composite blade made of two curvilinear reflectors (which I call sub-blades in the rest of the paper) rotating around a fixed axis. The incoming light beam is arriving parallel to the OX axis and it is displaced by $R_i$ along the OY direction. The beam is reflected by the rotating sub-blade number one onto the rotating sub-blade number two, which, finally, sends the beam out of the delay line. Independently of the rotation angle, the outgoing light beam follows the fixed line of path parallel to the OX axis; additionally, I suppose that the outgoing beam is displaced by $R_o$ along the OY axis. In what follows I demonstrate that an infinite number of solutions exist for such a composite blade, all of them parametrised by the choice of the coordinates of the two light reflection points at a given rotation angle. Moreover, the resulting optical delay of a composite blade system is always linear with the rotation angle.

I now consider a rotating blade that has two sub-blades featuring light reflection points defined by the coordinates $(x_i(\theta), R_i)$ and $(x_o(\theta), R_o)$. I use index $i$ to indicate the sub-blade that intercepts the incoming beam, while I use index $o$ to indicate the sub-blade that sends the beam out of the delay line. The corresponding blade inclination angles with respect to the OX axis are $\alpha_i(\theta)$ and $\alpha_o(\theta)$. As follows from Fig. 5(a), in order for the incoming and outcoming beams to follow two parallel paths I have to demand that for any rotation angle $\theta$ the inclination angles of the two blades are related as $\alpha_i(\theta) - \alpha_o(\theta) = \pi/2$. Using definition (6) I then get:

$$\begin{cases} \dfrac{x_o(\theta)}{\partial x_o(\theta)/\partial\theta - R_0} = \tan(\alpha_o(\theta)) \\ \dfrac{x_i(\theta)}{\partial x_i(\theta)/\partial\theta - R_i} = -\dfrac{1}{\tan(\alpha_o(\theta))} \end{cases} \Rightarrow \begin{cases} \dfrac{\partial x_o(\theta)}{\partial\theta} = R_0 + \dfrac{x_o(\theta)}{\tan(\alpha_o(\theta))} \\ \dfrac{\partial x_i(\theta)}{\partial\theta} = R_i - x_i(\theta)\tan(\alpha_o(\theta)) \end{cases} . \quad (39)$$

Moreover, as follows from Fig. 5(a), coordinates of the two light reflection points are related by the following geometrical relation:

$$x_i(\theta) - x_o(\theta) = (R_i - R_0)\tan\left(2\alpha_o - \dfrac{\pi}{2}\right) = \dfrac{R_i - R_0}{\tan(2\alpha_o)}. \quad (40)$$

By using the trigonometric identity $\tan(2\alpha) = 2\tan(\alpha)/(1-\tan^2(\alpha))$ I can now solve equation (40) with respect to the tangent of the angle $\alpha_o(\theta)$:

$$2\dfrac{x_i(\theta) - x_o(\theta)}{R_i - R_0} = \tan^{-1}(\alpha_o) - \tan(\alpha_o) \Rightarrow$$

$$\tan(\alpha_o(\theta)) = \dfrac{x_i(\theta) - x_o(\theta)}{R_i - R_0} \pm \sqrt{\left(\dfrac{x_i(\theta) - x_o(\theta)}{R_i - R_0}\right)^2 + 1} \quad . (41)$$

Finally, equations (39), (41) constitute a system of coupled differential equations that can be readily solved numerically. Noting that (39), (41) are the first order differential equations, therefore, solution for the blade shapes will be unique when specifying the positions of the two light reflection points at a certain rotation angle, say $x_i(\theta)\big|_{\theta=0} = x_i^0$, $x_o(\theta)\big|_{\theta=0} = x_o^0$.

In Figs. 5(b),(c), I present example of an optical delay line found by numerical solution of equations (39), (41) with initial conditions $x_i(0) = \sqrt{R_{dl}^2 - r_i^2}$, $x_o(0) = 0$.

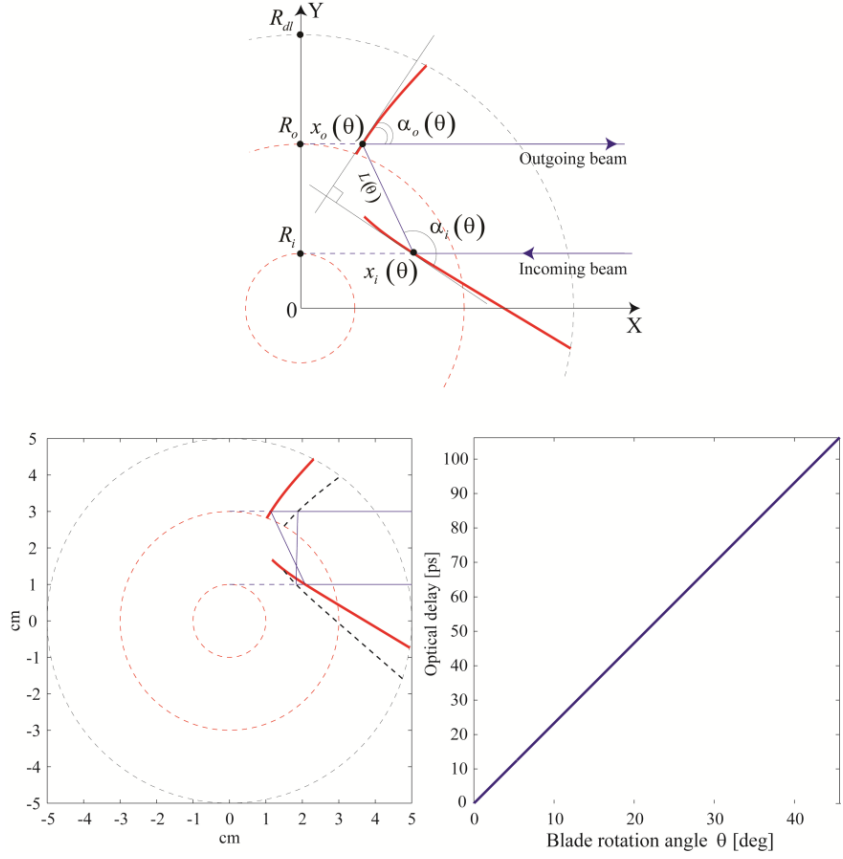

Fig. 5. (a) Schematic of a linear rotary optical delay line featuring a composite rotating blade with two sub-blades. (b) An example of a rotary delay line. Blade positions at various rotation angles are presented as dotted lines. (c) Corresponding optical delay.

### 4.1 Linear response of the rotary delay lines with two sub-blades

Remarkably, rotary delay lines that feature two sub-blades and offer collinear paths of the incoming and outgoing light beams show a linear optical delay that has a very simple form:

$$\Delta T = \frac{1}{c} \cdot \left( x_i(\theta) + x_o(\theta) - L(\theta) - \left( x_i(0) + x_o(0) - L(0) \right) \right) = \frac{R_i + R_o}{c} \theta \ , \quad (42)$$

where $L(\theta)$ is the distance between the two light reflection points given by:

$$L(\theta) = \sqrt{\left( x_i(\theta) - x_o(\theta) \right)^2 + \left( R_i - R_o \right)^2} = \frac{R_o - R_i}{\sin(2\alpha_o(\theta))} = -\frac{(R_i - R_o)}{2} \frac{1 + \tan^2(\alpha_o(\theta))}{\tan(\alpha_o(\theta))}$$

. (43)

Linear optical delay stated by (42) is relatively straightforward to confirm by taking a derivative of the optical delay with respect to the rotation angle and showing that this derivative takes a constant value. Particularly:

$$\frac{\partial (c \Delta T)}{\partial \theta} = \frac{\partial}{\partial \theta} \cdot \left( x_i(\theta) + x_o(\theta) - \sqrt{(x_i(\theta) - x_o(\theta))^2 + (R_i - R_o)^2} \right) =$$

$$\frac{\partial x_i(\theta)}{\partial \theta} + \frac{\partial x_o(\theta)}{\partial \theta} - \frac{(x_i(\theta) - x_o(\theta))\left(\frac{\partial x_i(\theta)}{\partial \theta} - \frac{\partial x_o(\theta)}{\partial \theta}\right)}{L(\theta)} = \qquad (44)$$

$$R_i + R_o + x_o \tan^{-1}(\alpha_0) - x_i \tan(\alpha_0) + 2 \frac{(x_i - x_o)(R_i - R_o - x_o \tan^{-1}(\alpha_0) - x_i \tan(\alpha_0))}{(R_i - R_o)(\tan^{-1}(\alpha_0) + \tan(\alpha_0))} =$$

$$R_i + R_o$$

where I have used (41) to simplify the expression.

### 5. Optical delay line using two rotating blades and a single stationary reflector

In this final section I consider another type of the optical delay line that can separate in space the incoming and outgoing light beams. It comprises a composite blade and a stationary reflector (see Fig. 6(a)). A composite blade is made of two curvilinear reflectors (sub-blades) rotating around a fixed axis. The incoming light beam is arriving parallel to the OX axis and it is displaced by $R_i$ along the OY axis. The beam is then reflected by the rotating sub-blade number one in the direction of a stationary reflector. Then, the light beam is redirected by the stationary reflector onto the rotating sub-blade number two, which, finally, sends the beam back out of the optical delay line. Independently of the rotation angle, the outgoing light beam follows the same line of path parallel to the OX axis which intersects the OY axis at $R_o$.

In the following analysis I distinguish two cases. Firstly, I do not impose linear response (with respect to the rotation angle) on the rotary delay line. In this case there exit an infinite number of solutions for the blade shapes. In fact, by arbitrarily choosing the shape of one of the sub-blades, and by specifying the position and inclination of the second sub-blade for the initial value of the rotation angle, then, the shape of the second sub-blade and the shape of the stationary reflector can be found by solving numerically a single nonlinear differential equation. Secondly, I impose linear response of the rotary delay line on the rotation angle. In this case, the blade shapes can be, in principal, found by solving a system of two nonlinear differential equations by specifying the positions and inclinations of the two blades for the initial value of the rotation angle.

I now consider a rotating blade that has two sub-blades featuring light reflection points defined by the coordinates $(x_i(\theta), R_i)$ and $(x_o(\theta), R_o)$. As before, I use index $i$ for the blade that intercepts the incoming beam, while I use index $o$ for the blade that sends the beam out of the delay line. The corresponding blade inclination angles with respect to the OX axis are $\alpha_i(\theta)$ and $\alpha_o(\theta)$. Finally, I define $\alpha_r(\theta)$ to be the inclination of the stationary reflector with respect to the OX axis at the reflection point that corresponds to the $\theta$ rotation angle of the delay line. As follows from Fig. 6(a), in order for the incoming and outcoming beams to follow two parallel paths I have to demand that for any rotation angle $\theta$ the inclination angles of the two blades are related to the reflector inclination angle as $\alpha_r(\theta) = \alpha_i(\theta) + \alpha_o(\theta) - \pi/2$. Using definition (6) I then get:

$$\tan(\alpha_r(\theta)) = -\cot(\alpha_i(\theta) + \alpha_o(\theta)) = \frac{1 - \cot(\alpha_i(\theta))\cot(\alpha_o(\theta))}{\cot(\alpha_i(\theta)) + \cot(\alpha_o(\theta))}$$

$$\cot(\alpha_o(\theta)) = \frac{\partial x_o(\theta)/\partial\theta - R_o}{x_o(\theta)} \quad ; \quad \cot(\alpha_i(\theta)) = \frac{\partial x_i(\theta)/\partial\theta - R_i}{x_i(\theta)}$$

. (45)

I now denote $L_i(\theta)$ and $L_o(\theta)$ to be the distances from the light reflection points on the blade surfaces to the light reflection point on the surface of a stationary reflector. Using the Cartesian coordinates of the light reflection points on the blade surfaces $(x_i(\theta), R_i)$, and $(x_o(\theta), R_o)$, I now find the coordinates $(x_r(\theta), y_r(\theta))$ of the light reflection point on the surface of a stationary reflector. From geometrical considerations (see Fig. 6(a)) I write:

$$x_r(\theta) = x_o(\theta) + L_o(\theta)\cos(\pi - 2\alpha_o(\theta)) = x_i(\theta) + L_i(\theta)\cos(2\alpha_i(\theta) - \pi)$$

$$y_r(\theta) = R_o - L_o(\theta)\sin(\pi - 2\alpha_o(\theta)) = R_i + L_i(\theta)\sin(2\alpha_i(\theta) - \pi)$$

, (46)

from which it follows that:

$$L_o(\theta) = \frac{(x_i(\theta) - x_o(\theta))\tan(2\alpha_i(\theta)) - (R_i - R_o)}{(\tan(2\alpha_o(\theta)) - \tan(2\alpha_i(\theta)))\cos(2\alpha_o(\theta))}$$

$$L_i(\theta) = \frac{(x_i(\theta) - x_o(\theta))\tan(2\alpha_o(\theta)) - (R_i - R_o)}{(\tan(2\alpha_o(\theta)) - \tan(2\alpha_i(\theta)))\cos(2\alpha_i(\theta))}$$

. (47)

$$x_r(\theta) = \frac{x_o(\theta)\tan(2\alpha_o(\theta)) - x_i(\theta)\tan(2\alpha_i(\theta)) + (R_i - R_o)}{\tan(2\alpha_o(\theta)) - \tan(2\alpha_i(\theta))}$$

$$y_r(\theta) = \frac{R_i\cot(2\alpha_i(\theta)) - R_o\cot(2\alpha_o(\theta)) - (x_i(\theta) - x_o(\theta))}{\cot(2\alpha_i(\theta)) - \cot(2\alpha_o(\theta))}$$

Finally, I need to demand that the tangent to the reflector shape given by (47) equals to (45):

$$\frac{\partial y_r(\theta)/\partial\theta}{\partial x_r(\theta)/\partial\theta} = \tan(\alpha_r(\theta)), (48)$$

which results in the differential equation that relates two blade shapes $x_i(\theta)$ and $x_o(\theta)$. Given the shape of one sub-blade, say $x_o(\theta)$, then equation (48) can be solved numerically for the shape of the second sub-blade $x_i(\theta)$ if the starting point $x_i(0)$ and $\partial x_i(\theta)/\partial\theta|_{\theta=0}$ of the second sub-blade are known at a given value of the rotation angle. Note also that all the expressions in (47) and (48) can be readily evaluated after presenting the trigonometric functions of a double argument in terms of the tangents of a single argument that can be calculated from $x_i(\theta)$, $x_o(\theta)$ using (45). Namely, the following trigonometric identities can be used:

$$\tan(2\alpha) = 2\tan(\alpha)/(1-\tan^2(\alpha))$$
$$\cos(2\alpha) = (1-\tan^2(\alpha))/(1+\tan^2(\alpha))$$
. (49)

### 5.1 Linear delay line

In order to design optical delay lines with linear response, in addition to condition (48) I have to demand that:

$$\Delta T(\theta) = \frac{1}{c} \cdot (x_i(\theta) + x_o(\theta) - L_i(\theta) - L_o(\theta) - (x_i(0) + x_o(0) - L_i(0) - L_o(0))) = \beta \cdot \theta$$
, (50)

where $\beta$ is a constant. By taking a derivative of (50) with respect to the rotation angle, I finally arrive to a system of two coupled differential equations:

$$\begin{cases} \dfrac{\partial y_r(\theta)/\partial \theta}{\partial x_r(\theta)/\partial \theta} - \tan(\alpha_r(\theta)) = 0 \\ \dfrac{\partial \Delta T(\theta)}{\partial \theta} - \beta = 0 \end{cases}$$
, (51)

which can be, in principle, resolved numerically with respect to the unknown blade shapes $x_i(\theta)$ and $x_o(\theta)$. In order to find a numerical solution of (51) one has to specify the starting points $x_o(0)$, $x_i(0)$, and derivatives $\partial x_o(\theta)/\partial \theta\big|_{\theta=0}$, $\partial x_i(\theta)/\partial \theta\big|_{\theta=0}$ for the two blades at a given value of the rotation angle.

### 5.2 Iterative method for finding the blade shapes

In this final subsection I present example of an iterative algorithm that can be used to find numerically the shape of one sub-blade (say $x_i(\theta)$) assuming that the shape of the other sub-blade (say $x_o(\theta)$) is known. This amounts to numerical solution of equation (48).

First I assume that at a given value of the rotation angle $\theta_0$ the values of $x_{o,i}(\theta_0)$, $\partial x_{i,o}(\theta)/\partial \theta\big|_{\theta=\theta_0}$, and $\partial^2 x_o(\theta)/\partial \theta^2\big|_{\theta=\theta_0}$ are known. Then, equation (48) can be solved numerically with respect to the second order derivative $\partial^2 x_i(\theta)/\partial \theta^2\big|_{\theta=\theta_0}$. Particularly I note that (48) can be written in terms of the partial derivatives with respect to the functions $x_{i,o}(\theta)$ and their derivatives:

$$\frac{\partial y_r(\theta)/\partial \theta}{\partial x_r(\theta)/\partial \theta}\bigg|_{\theta=\theta_0} = \frac{\dfrac{\partial y_r}{\partial x_o}x'_o + \dfrac{\partial y_r}{\partial x'_o}x''_o + \dfrac{\partial y_r}{\partial x_i}x'_i + \dfrac{\partial y_r}{\partial x'_i}x''_i}{\dfrac{\partial x_r}{\partial x_o}x'_o + \dfrac{\partial x_r}{\partial x'_o}x''_o + \dfrac{\partial x_r}{\partial x_i}x'_i + \dfrac{\partial x_r}{\partial x'_i}x''_i} = \tan(\alpha_r), (52)$$

where to simplify the notation I have used $x_{i,o} = x_{i,o}(\theta_0)$, $x'_{i,o} = \partial x_{i,o}(\theta)/\partial \theta\big|_{\theta=\theta_0}$, $\alpha_r = \alpha_r(\theta_0)$. Als,o note that all the partial derivatives in (52) can be easily evaluated as $x_r$

and $y_r$ have explicit dependence on $x_{i,o}$ and $x'_{i,o}$ as follows from (45) and (47). By noticing that none of the partial derivatives with respect to $x_{o,i}$ or $x'_{o,i}$ depend on $x''_i$, I can then resolve (52) as:

$$\left.\frac{\partial^2 x_i(\theta)}{\partial \theta^2}\right|_{\theta=\theta_0} = \frac{\left(\tan(\alpha_r)\frac{\partial x_r}{\partial x_o} - \frac{\partial y_r}{\partial x_o}\right)x'_o + \left(\tan(\alpha_r)\frac{\partial x_r}{\partial x'_o} - \frac{\partial y_r}{\partial x'_o}\right)x''_o + \left(\tan(\alpha_r) - \frac{\partial y_r}{\partial x_i}\right)\frac{\partial x_r}{\partial x_i}x'_i}{\left(\frac{\partial y_r}{\partial x'_i} - \tan(\alpha_r)\frac{\partial x_r}{\partial x'_i}\right)}$$

. (53)

Finally, for the values of $x_i(\theta_0 + d\theta)$ and $\left.\partial x_i(\theta)/\partial \theta\right|_{\theta=\theta_0 + d\theta}$ I write:

$$x_i(\theta_0 + d\theta) = x_i(\theta_0) + d\theta \cdot \left.\frac{\partial x_i(\theta)}{\partial \theta}\right|_{\theta=\theta_0} + \frac{d\theta^2}{2} \cdot \left.\frac{\partial^2 x_i(\theta)}{\partial \theta^2}\right|_{\theta=\theta_0}$$

$$\left.\frac{\partial x_i(\theta)}{\partial \theta}\right|_{\theta=\theta_0 + d\theta} = \left.\frac{\partial x_i(\theta)}{\partial \theta}\right|_{\theta=\theta_0} + d\theta \cdot \left.\frac{\partial^2 x_i(\theta)}{\partial \theta^2}\right|_{\theta=\theta_0}$$

. (54)

Iterations (53), (54) can then be repeated multiple time to find the function $x_i(\theta)$.

In Figs. 6(b),(c) I present example of a rotary optical delay line found by solving equation (48) using iterative algorithm presented above. In this case I have used $x_o(\theta) = (0.8 + 1.1 \cdot \theta) \cdot r_o$, $x_i(0) = r_i$, $\left.\partial x_i(\theta)/\partial \theta\right|_{\theta=0} = 0.95 \cdot r_i$, $r_i = 1\ cm$, $r_o = 3\ cm$. The parameters were chosen to have a stationary reflector positioned outside of the rotating part of a delay line. As in all the previous examples considered in this work, the blades are inscribed into a $10\ cm$-diameter circle. For this design, the angle dependence of the optical delay is virtually linear and it provides up to $130\ ps$ optical delay. The usable range of rotation angles is $\approx 28°$, therefore, up to 13 reflectors can be integrated into a single delay line, thus resulting in an order of magnitude enhancement of the acquisition rate compared to a delay line with a single reflector.

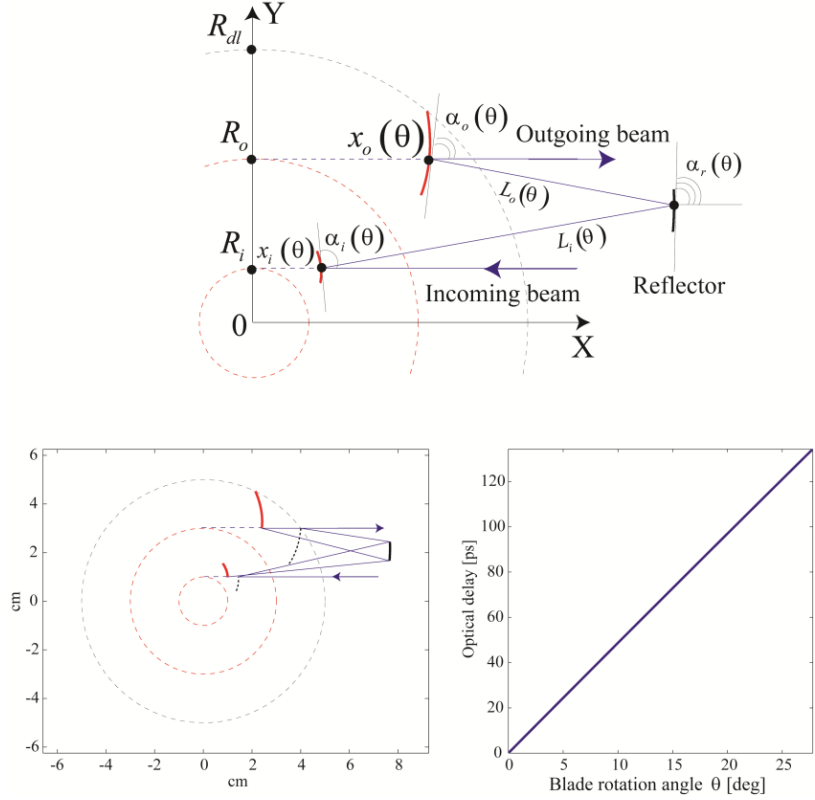

Fig. 6. (a) Schematic of a rotary optical delay line featuring a composite rotating blade with two sub-blades and a stationary reflector. (b) An example of a rotary delay line. Blade positions at various rotation angles are presented as dotted lines. (c) Corresponding optical delay.

## 6. Examples of practical realisation of the curvilinear reflectors using CNC machining.

In this final sections I discuss two prototypes of the rotary delay lines that were fabricated using CNC machining. The goal of this section is only to show that designs presented in this papers are experimentally feasible considering current advances in the computer-controlled micro-machining. As this paper is mostly theoretical, I defer the detailed discussion of the optical characterisation of the presented prototypes to our future publications.

The two prototypes were fabricated from the 5mm-thick acrylonitrile butadiene styrene (ABS) sheets using a Charly 2U CNC machine that offers precision of linear translation of 5 μm. A self-adhesive reflective aluminium-foil of ~50 μm thickness was attached to the machined blade surface to form curvilinear reflectors.

The first prototype (see Fig. 7(a)) comprises a single rotating planar reflector and a single stationary curvilinear reflector as shown in Fig. 4(a). In this arrangement, the incoming and the outgoing light passes coincide in space. A stationary reflector is machined together with a small circular guide for the rotary reflector in order to simplify alignment of the two parts. Rotating reflector was mounted on the multi-axis optomechanic stage in order to control the blade spatial position, as well as the blade rotation angle. The rotating reflector was inscribed into the disc of radius $R_{dl} = 5\,cm$. Shape of the stationary curvilinear reflector was calculated as described in section 3 and (33) with $R_i = R_{dl} \cdot \cos(2\pi/7) \approx 3.12\,cm$, $\beta = 2R_i$, $L_0 = 2R_{dl}$, $\theta_0 = 0$. By varying the blade rotation angle between 0 and 35 deg, an optical delay of ~250ps was measured experimentally directly from the high-resolution images of the light path. As seen in Fig 7(b), optical delay is linearly dependent on the rotation angle, which is in accordance with the theoretical prediction. Note that precision of this measurement is limited to ~6ps, which is due to low precision (~2mm) associated with the direct measurement of the optical light path.

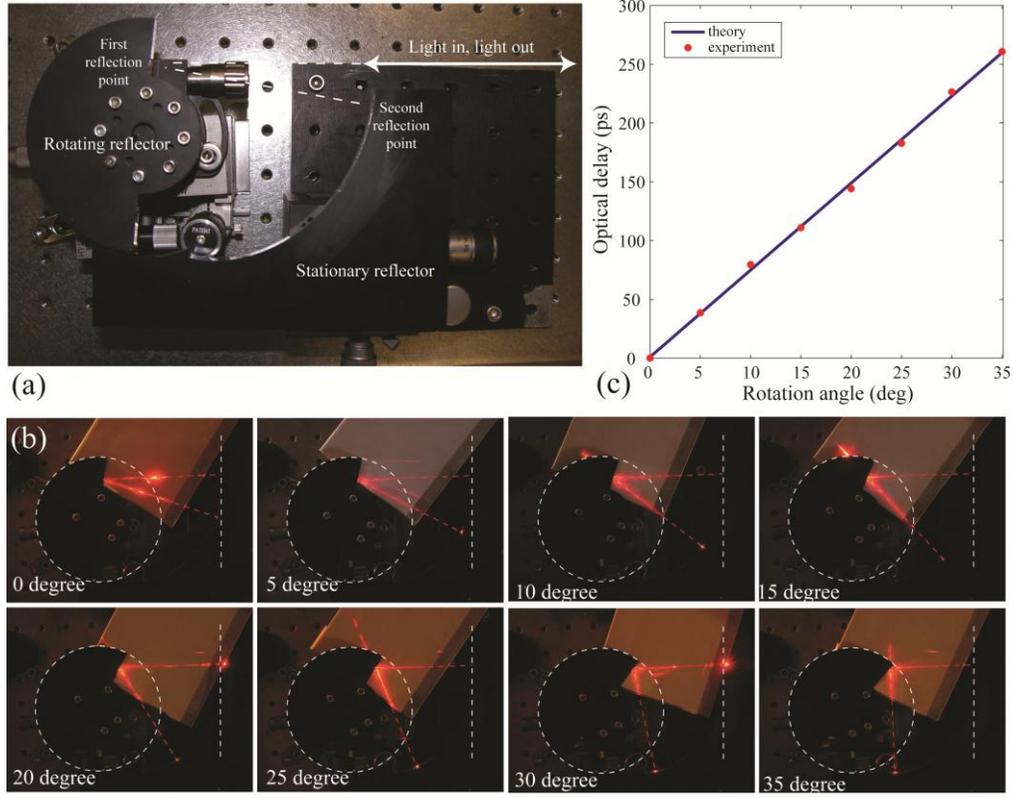

Fig. 7. (a) Experimental realisation of the rotary delay line featuring one rotating and one stationary reflector. (b) Optical delay measured directly from the high-resolution images of the light path. (c) Optical delay.

The second prototype (see Fig. 8(a)) comprises two rotating curvilinear reflectors designed in such a way that the outgoing and incoming light patch are separated in space. The rotating reflector was inscribed into the disc of radius $R_{dl} = 5\,cm$. Shape of the stationary curvilinear reflector was calculated as detailed in section 4 and using $x_i(0) = \sqrt{R_{dl}^2 - R_i^2}$, $x_o(0) = 0$, where $R_{dl} = 5\,cm$, $R_i = 2.97\,cm$, $R_o = 3.8\,cm$. By varying the blade rotation angle between 0 and 35 deg, an optical delay of ~120ps was experimentally measured from the images of the optical light path. As seen in Fig 8(b), optical delay is linearly dependent on the rotation angle, which is in accordance with the theoretical prediction.

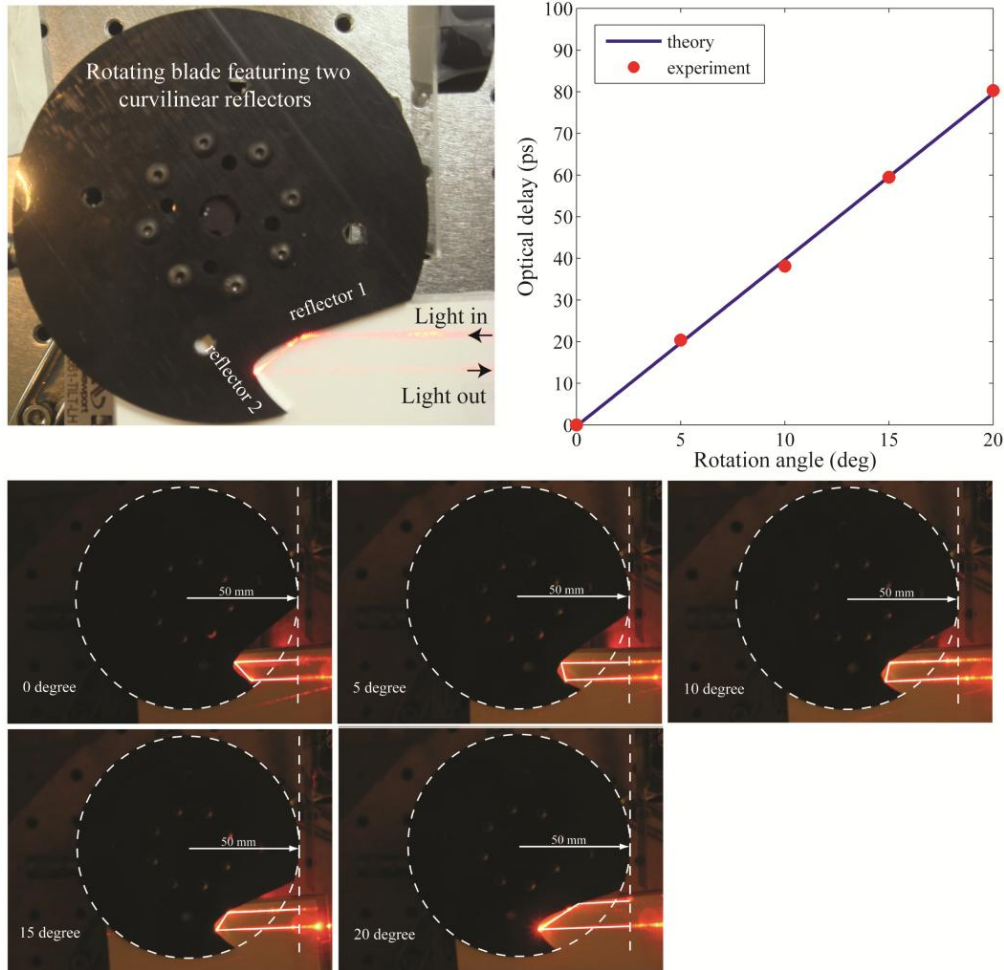

Fig. 8. (a) Experimental realisation of the rotary delay line featuring one rotating and one stationary reflector. (b) Optical delay measured directly from the high-resolution images of the light path. (c) Optical delay.

**Conclusions**

In this work I have developed a comprehensive theory for the design of rotary delay lines that feature a combination of rotating and stationary reflectors. Particularly, I have presented theoretical formulations as well as analytical and semi-analytical solutions for four general classes of the rotary optical delay lines. Delay lines under consideration are comprised of the following components: a single rotating reflector, a single rotating reflector and a single stationary reflector, two rotating reflectors, and, finally, two rotating reflectors and a single stationary reflector. I have demonstrated that in all these cases it is possible to design a variety of linear optical delay lines via optimization of shapes of the rotating and stationary reflector surfaces. Moreover, in the case of two rotating reflectors a convenient spatial separation of the incoming and outgoing beams is possible. In the case of a single rotating reflector spatial separation of the incoming and outgoing beams is also possible via introduction of the rectangular V groove into the reflector edge. For the sake of example, all the blades presented in this paper were chosen to fit into a circle of 10cm diameter and these delay lines feature in excess of 600ps of optical delay. This is comparable to the optical delay generated by a standard 10cm-long linear micropositioning stage with a retroreflector, while in addition, rotary delay lines offer high acquisition rates of at least several 100s Hz.

**Acknowledgements**
This work was supported by the NSERC Strategic Grant entitled "A dynamically reconfigurable THz-TDS imaging system for industrial measurement applications". I thank Dr. Hang Qu for his contribution in the experimental characterization of the optical delay for the two prototypes presented in section 6 of this paper.

**Appendix**

Here I derive expression for the inclination angle of the blade surface with respect to the direction of incoming light (OX axis) at the point of light reflection. In order to do that I consider the blade position at two values of the rotation angle $\theta$ and $\theta + \delta\theta$. Particularly, for the blade rotation angle $\theta$ the position of the light reflection point is given by $(x_b(\theta), R_i)$. After blade rotation by $\delta\theta$, this point will have the new coordinates given by:

$$\begin{pmatrix} x_1 \\ y_1 \end{pmatrix} = \begin{pmatrix} \cos(\delta\theta) & \sin(\delta\theta) \\ -\sin(\delta\theta) & \cos(\delta\theta) \end{pmatrix} \begin{pmatrix} x_b(\theta) \\ R_i \end{pmatrix} \approx \begin{pmatrix} x_b(\theta) + R_i \delta\theta \\ R_i - x_b(\theta)\delta\theta \end{pmatrix} + O(\delta\theta^2). \quad \text{(A.1)}$$

At the same time, coordinates of the new light reflection point will be $(x_2, y_2) = (x_b(\theta + \delta\theta), R_i)$, which allows us to find the inclination of the blade with respect to the OX axis at the light reflection point as:

$$\tan(\alpha_b) = \lim_{\delta\theta \to 0} \frac{y_2 - y_1}{x_2 - x_1} = \frac{x_b(\theta)}{\partial x_b(\theta)/\partial\theta - R_i}. \quad \text{(A.2)}$$